\documentclass[12pt,noshowpacs,nofootinbib,notitlepage,amsmath,amssymb]{revtex4-2}
\usepackage{setspace}
\allowdisplaybreaks
\usepackage{graphicx,color}
\usepackage[colorlinks=true,citecolor=blue,linkcolor=blue,urlcolor=blue]{hyperref}
\usepackage[charter]{mathdesign}
\usepackage{multirow}
\usepackage{enumerate}
\usepackage{array}
\usepackage{booktabs}

\DeclareSymbolFontAlphabet{\mathcal}{symbols}
\DeclareSymbolFont{symbols}{OMS}{xmdcmsy}{m}{n}
\DeclareSymbolFont{largesymbols}{OMX}{cmex}{m}{n}
\SetSymbolFont{symbols}{bold}{OMS}{xmdcmsy}{b}{n}

\begin{document}  
\title{\color{blue}\Large 2-2-holes simplified}

\author{Bob Holdom}
\email{bob.holdom@utoronto.ca}
\affiliation{Department of Physics, University of Toronto, Toronto, Ontario, Canada  M5S 1A7}
\begin{abstract}
Quadratic gravity illustrates how a replacement for black holes can emerge from a UV completion of gravity. 2-2-holes are extremely compact horizonless objects with an entropy $S_{22}$ due to trapped normal matter, and in this way they are conceptually easy to understand. But the field equations are cumbersome and the numerical analysis has so far been restricted to relatively small size solutions. Here we show how the properties of arbitrarily large 2-2-holes can be found, including the time delay for gravitational wave echoes and the result $T_\infty S_{22}=M/2$. The starting point is to formulate the metric in terms of the tortoise coordinate, and to have one of the two metric functions be a conformal factor. A large conformally-related volume becomes associated with the interior of a 2-2-hole. We also discuss implications for the weak gravity conjecture.
\end{abstract}

\maketitle 

\section{Introduction}
A 2-2-hole is a horizonless self-gravitating solution of quadratic gravity that is sourced by normal matter \cite{Holdom:2016nek,Holdom:2019ouz,Ren:2019jft}. It reproduces the Schwarzschild solution down to radii of order a proper Planck length from the would-be horizon. Its interior is characterized by large curvature and small proper volume. In \cite{Holdom:2016nek} matter sources in the form of spherical shells were studied. As the shell radius decreases and approaches the Schwarzschild radius from above, curvature invariants such as the Weyl-squared invariant grow large for radii in the vicinity of the shell. This does not happen in general relativity because of Birkhoff’s theorem, but this theorem does not apply to quadratic gravity. When the shell radius becomes smaller than the Schwarzschild radius, it might then be expected that the large curvatures persist by continuity. This is the case for a 2-2-hole where the effects of strong gravity continue to extend up to and very slightly beyond the would-be horizon no matter where the inner matter shell is located. This suggests that if quadratic gravity is the UV completion of general relativity then it is 2-2-holes rather than black holes that are the endpoint of gravitational collapse.

A more realistic matter source is a relativistic thermal gas with $T^\mu_\mu=0$, i.e.~$\rho=3p$, as considered in \cite{Holdom:2019ouz,Ren:2019jft}. For either matter source the interior solution displays a simple scaling law with respect to the 2-2-hole mass $M$. For the thermal-gas 2-2-hole this scaling law implies that the total entropy satisfies an area law that is similar in size to the entropy of the same mass black hole. We shall review this more below.

2-2-holes and black hole exteriors may be compared in terms of either of the two line elements
\begin{align}
ds^2&=-B(r)dt^2+A(r)dr^2+r^2 d\theta^2+r^2 \sin(\theta)^2 d\phi^2,\label{e16}\\
&=-\tilde B(x)dt^2+\tilde B(x)dx^2+R(x)^2 d\theta^2+R(x)^2 \sin(\theta)^2 d\phi^2\label{e19}
.\end{align}
The second form introduces the tortoise coordinate $x$ such that $g_{tt}=-g_{xx}$, and so $x$ is defined by $dx/dr=\sqrt{A(r)/B(r)}$ with $R(x)=r$ and $\tilde B(x)=B(R(x))$. The black hole exterior is characterized by a horizon located at $x=-\infty$, whereas the 2-2-hole has no horizon and instead the origin of the 2-2-hole is located at a finite tortoise coordinate. Thus an observer outside a 2-2-hole may see a signal reach the origin and return in a finite time. This is the most obvious physical characteristic that distinguishes 2-2-holes from black holes.

We also note that $\tilde B(x)=0$ occurs at the origin of the 2-2-hole, as it does at the horizon of the black hole. Thus in a sense the horizon of a black hole is pushed into the single point at the origin of the 2-2-hole. The whole 2-2-hole metric vanishes at the origin and the result is a naked time-like curvature singularity. This is a benign singularity with interesting physical consequences, as discussed in \cite{Holdom:2016nek,Holdom:2019ouz,Ren:2019jft}, and as we shall review below.

Quadratic gravity offers a UV completion of gravity in the form of a quantum field theory. This QFT is defined by the action
\begin{align}
S&=\frac{1}{16\pi}\int d^4x \sqrt{-g}\left(m_{\rm Pl}^2R- \frac{1}{2}\frac{m_{\rm Pl}^2}{m_G^2}C^{\mu\nu\alpha\beta}C_{\mu\nu\alpha\beta} + \frac{1}{6}\frac{m_{\rm Pl}^2}{m_S^2}R^2\right)
,\label{e8}\end{align}
where $m_G$ ($m_S$) is the ghost mass (gravi-scalar mass). Due to renormalization effects, $m_G/m_{\rm Pl}$ becomes an asymptotically-free running coupling. The QFT is stable and unitary and it is only in this context that the massive ghost state can be properly treated and understood \cite{Donoghue:2019ecz,Donoghue:2019fcb} (see also \cite{Holdom:2019ouz}). The perturbative QFT can then be used to study ultra-Planckian scattering \cite{Holdom:2021hlo,Holdom:2021oii}.

Our interest here is with the macroscopically large states of this gravity theory, and for this we turn to the static solutions of the classical field equations. There would normally be the question of why a classical theory is being considered with two and four derivatives and not higher derivatives. This receives some explanation in the context of the renormalizable and asymptotically-free QFT that is defined by the $2+4$ derivative action. In particular the high curvatures inside the 2-2-hole are probing the asymptotically free region of the QFT. The singularity itself has a description that lies within the UV complete theory.

Our study of the thermal-gas 2-2-hole solutions shall take place for the most part in the classical Einstein-Weyl theory. This corresponds to the action in the limit $m_S\to\infty$, that is with no $R^2$ term. This removes the gravi-scalar degree of freedom, and it simplifies the analysis considerably. The Einstein-Weyl theory has the remaining parameter $m_G$ that we might expect to be of order $m_{\rm Pl}$. The thermal-gas solutions are not specific to the Einstein-Weyl theory, and in fact we shall find that they are also solutions of the full theory for any $m_S$.

Previous studies of 2-2-holes have used the metric in (\ref{e16}), which has a couple of advantages. The field equations are independent of the overall scale of $B(r)$, which can then be simply fixed from the boundary condition $B(\infty)=1$. And in the case of the Einstein-Weyl theory the field equations can be reduced to be no more than second order in derivatives \cite{Lu:2015psa}. But there is also a disadvantage, and that is that the 2-2-hole solution has $A(r)$ and $B(r)$ varying very rapidly for $r\approx 2M$. $A(r)$ in particular has an extremely sharp peak that grows for increasing $M$, due to how close the exterior solution is to the Schwarzschild solution. This makes it numerically very demanding to study solutions with $M$ much greater than a few hundred in Planck units. This has been sufficient to find the scaling law for the interior region that is inside of the $A(r)$ peak location. But this leaves an important physical property of realistically large 2-2-holes quite uncertain. This is the time delay for gravitational wave echoes, that is the travel time for a wave to travel from the light-ring to the origin and back out again. The wave lingers near the peak region where the apparent speed of light drastically drops.

We shall make progress by choosing a different form for the general spherically symmetric metric. There are again two metric functions, but these are much smoother in character. The interior part of the 2-2-hole solution very smoothly matches onto the black hole solution of the exterior. (Black holes in this paper shall always refer to the Schwarzschild solution, and not the non-Schwarzschild black hole solutions that also exist in quadratic gravity \cite{Lu:2015psa,Svarc:2018coe}.) Given this, a series expansion around the origin becomes more useful and the numerical analysis is easier. We are then able to obtain all the desired physical characteristics of arbitrarily large 2-2-holes, including the time delay.

The form of the metric we choose is also of interest more generally. Basic results such as wave equations and geodesic equations take a particularly simple form in terms of the metric functions. The field equations become simple enough to look at and to understand how the 2-2-hole solution is emerging. The metric also points to a very large conformally-related volume that emerges in the calculation of the 2-2-hole entropy, as discussed in Section \ref{s1}. Sections \ref{s4} and \ref{s5} start with the field equations and end up with the properties of large 2-2-holes. Section \ref{s2} is devoted to the wave equation and gravitational waves in particular. Section \ref{s3} considers how 2-2-holes behave under the conditions that are used to justify the weak gravity conjecture in the case of black holes.

\section{Basic picture}
We describe the general spherically symmetric static spacetime in terms of the tortoise coordinate as in (\ref{e19}), but with with a new function $S(x)\equiv R(x)/\sqrt{\tilde B(x)}$,
\begin{align}
ds^2=-\frac{R(x)^2}{S(x)^2}dt^2+\frac{R(x)^2}{S(x)^2}dx^2+R(x)^2 d\theta^2+R(x)^2 \sin(\theta)^2 d\phi^2
.\label{e0}\end{align}
The two functions $R(x)$ and $S(x)$ are both length scales. $R(x)$ appears as a radius of spherical surfaces as a function of the tortoise coordinate. But $R(x)^2$ also appears as a conformal factor.\footnote{A metric with an overall conformal factor was also found to be useful for the study of non-Schwarzschild black-hole solutions in quadratic gravity in \cite{Svarc:2018coe}. In that case a non-diagonal metric was used.} The field equations of interest to us are not conformally invariant, and so this conformal factor cannot be removed by a conformal transformation. On the other hand we can consider other physics on top of the background spacetime. If this physics is conformally invariant then it will have a description that is independent of $R(x)$. For that physics we can make a conformal transformation and choose to put the line element into the form
\begin{align}
ds^2=-dt^2+dx^2+S(x)^2 d\theta^2+S(x)^2 \sin(\theta)^2 d\phi^2
.\label{e00}\end{align}
$S(x)$ provides the radii of another set of spherical surfaces that will be of interest, and we shall refer to (\ref{e00}) as the conformally-related metric.

We first consider the black hole, where the tortoise coordinate is related to the standard radial coordinate $r$ of the Schwarzschild metric by  
\begin{align}
x=r+2M\log(\frac{r-2M}{2M})+(2\log(2)-3)M+x_{\rm lr}
.\label{e14}\end{align}
We have added some constants and the significance of $x_{\rm lr}$ will become clear shortly. The range $-\infty<x<\infty$ corresponds to $2M<r<\infty$. For an observer at infinity, $\Delta x$ gives the light travel time to cover a radial distance $\Delta r$. The function $R(x)$ for a black hole is obtained by inverting (\ref{e14}) for $r=R_{\rm bh}(x)$, which can be done explicitly.\footnote{This was noticed for example in \cite{Boonserm:2008zg} and it can be done in \textit{Maple}.} Thus the black hole is represented by the metric in (\ref{e0}) when
\begin{align}
R_{\rm bh}(x)&=2M\left[1+W\!\!\left(\frac{1}{2}\exp(\frac{x-x_{\rm lr}+M}{2M})\right)\right],\\
S_{\rm bh}(x)&=\frac{R_{\rm bh}(x)}{\sqrt{1-2M/R_{\rm bh}(x)}}
\label{e4}.\end{align}
$W$ is the Lambert $W$ function, also known as the product logarithm, where $w=W(x)$ is defined by $w e^w=x$. $S_{\rm bh}(x)$ has its minimum at $x_{\rm lr}$ where $S_{\rm bh}(x_{
\rm lr})=3\sqrt{3}M$ and $R_{\rm bh}(x_{\rm lr})=3M$. $x_{\rm lr}$ thus corresponds to the light-ring radius.

The large $x$ behavior is obtained from $W(x)\sim \log(x)-\log(\log(x))$ for large $x$,
\begin{align}
R_{\rm bh}(x)&\to x-2M\log(\frac{x}{2M})-x_R(M)+{\cal O}(\frac{1}{x})\quad\mbox{and}\nonumber\\
S_{\rm bh}(x)&\to x-2M\log(\frac{x}{2M})-x_S(M)+{\cal O}(\frac{1}{x})\quad\mbox{for}\quad x-x_{\rm lr}\gg 2M
.\end{align}
We are more interested in the small $x$ behavior, as obtained from $W(x)\sim x$ for small $x$,
\begin{align}
R_{\rm bh}(x)&\to2M+M\exp(\frac{x-x_{\rm lr}+M}{2M})\quad\mbox{and}\nonumber\\
S_{\rm bh}(x)&\to 2\sqrt{2}M\exp(-\frac{x-x_{\rm lr}+M}{4M})\quad\mbox{for}\quad x-x_{\rm lr}\ll -2M
.\label{e11}\end{align}
Here $S_{\rm bh}(x)$ is growing exponentially for decreasing $x$ while $R_{\rm bh}(x)\to 2M$. Thus for a given $\Delta x$ the proper radial distance is becoming exponentially small, just as the proper time does for a given $\Delta t$.

\begin{figure}[h]
\includegraphics[width=.8 \textwidth]{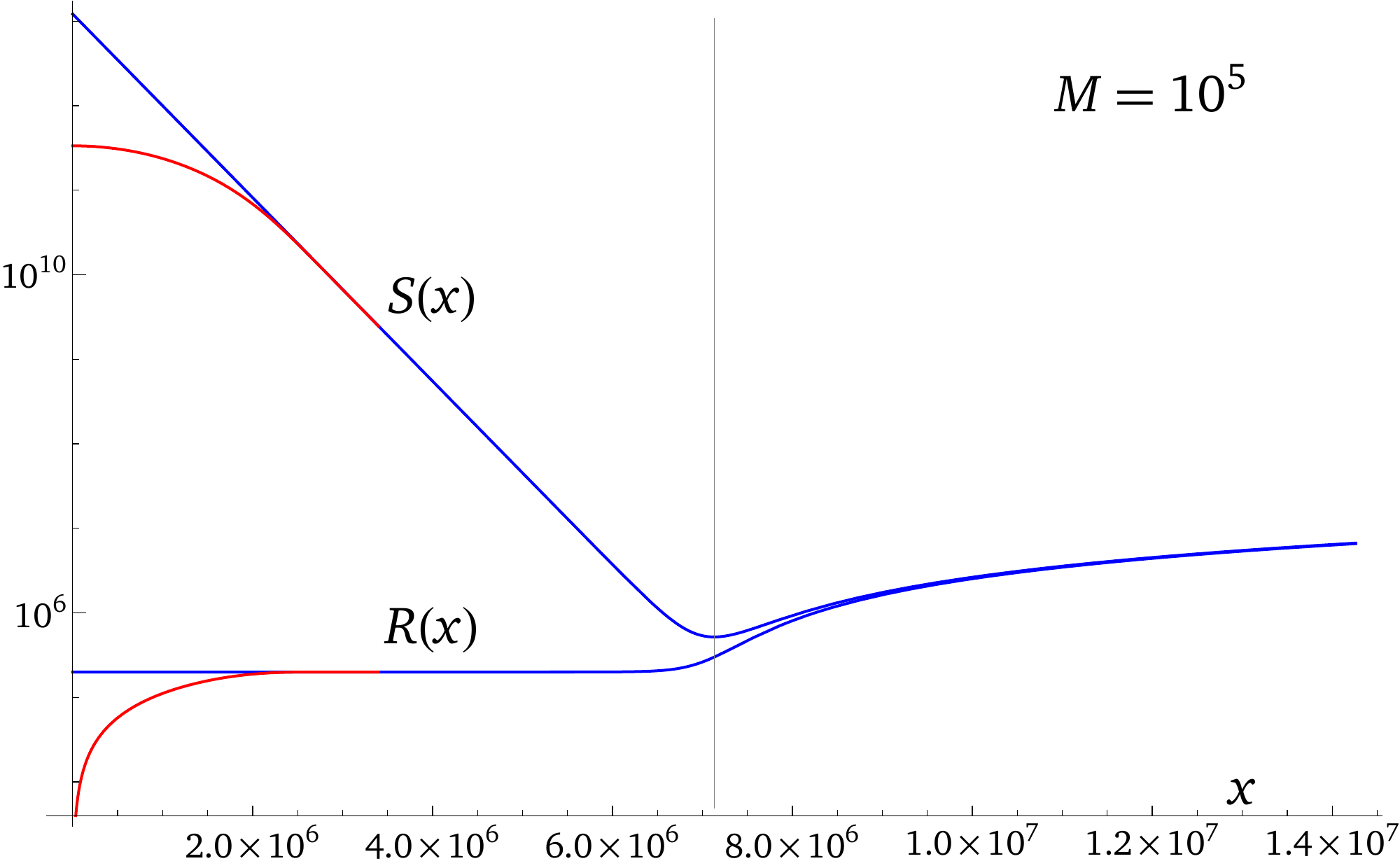}
\caption{The Schwarzschild solution in blue and the deviation caused by the 2-2-hole solution in red for $M=10^5$ in Planck units. Note that this is a log plot and $R(x)\sim x$ near the origin of a 2-2-hole. The vertical line at $x_{\rm lr}$ corresponds to the light-ring radius.}\label{f1}
\end{figure}

A 2-2-hole changes this black hole picture by providing a lower bound on $x$ where $R(x)$ goes to zero. This point is at the origin of the 2-2-hole, and this $x$ can be defined to be $x=0$. Then each value of $x> 0$ gives the light travel time from the radius $R(x)$ to the origin. Once the interior 2-2-hole solution is known and the exterior black hole solution is properly matched onto it, then the value of $x_{\rm lr}$ in the latter solution is also fixed. In Fig.~\ref{f1} we show in blue the black hole version of $R(x)$ and $S(x)$ for a finite range of $x$, and with $M=10^5$ in units with $m_{\rm Pl}=1$. These curves are modified by a 2-2-hole of the same mass, as shown by the red curves. Thus we see that a 2-2-hole basically puts a cap on the exponential growth of $S(x)$. We also see that the region where $R(x)\approx2M$ is quite featureless, with no obvious $x$ to associate with the 2-2-hole surface. In Sections \ref{s4} and \ref{s5} we discuss in detail how the red curves are obtained.

\section{The trapped matter}\label{s1}
In this section we recast various results in terms of the new metric functions $R(x)$ and $S(x)$. We first reconsider the calculation of the total entropy and energy of the relativistic thermal gas contained by a 2-2-hole. The associated densities in terms of the temperature $T(x)$ are
\begin{align}
s(x)=\frac{2\pi^2}{45}N T(x)^3,\quad \rho(x)=\frac{3}{4}T(x)s(x)=3p(x)
.\end{align}
$N$ accounts for the number of particle degrees of freedom, in units of the contribution of one massless boson. The total entropy $S_{22}$ is obtained by integrating over the volume element $\sqrt{-\tilde g}dx d\theta d\phi$ where $\tilde g$ is just the spatial part of the metric. The integral for the total matter energy $U_{22}$ uses $\sqrt{-g}$ instead. Meanwhile $T(x)$ is given by $T(x)=T_\infty /\sqrt{-g_{tt}}$, as deduced from $\nabla^\mu T_{\mu\nu}=0$. The metric (\ref{e0}) then gives the results
\begin{align}
S_{22}&=\frac{2\pi^2}{45}N T_\infty^3 V_{22},\quad U_{22}=\frac{3}{4}T_\infty S_{22}=3p_\infty V_{22},\nonumber\\
V_{22}&=4\pi\int_0^{x_{\rm lr}} S(x)^2 dx
\label{e15}.\end{align}

There is no dependence on $R(x)$ and so this is a situation where we can choose to use the conformally-related metric in (\ref{e00}). The volume $V_{22}$ is nothing more than the spatial volume as specified by that metric. For decreasing $x$, spherical surfaces have growing radii $S(x)$ until ending on the largest spherical surface at $x=0$. Thus we have an alternative prescription to calculate $S_{22}$ or $U_{22}$. Simply use the conformally-related metric to calculate a volume and assume a gas of constant temperature $T_\infty$ spread throughout this volume. This is self-consistent since $\nabla^\mu T_{\mu\nu}=0$ for this metric implies that $p'(x)=0$ for the pressure of the gas.

We shall find that $S(x)\sim M^2$ in the interior of a 2-2-hole and thus $V_{22}\sim M^5$. Then with $T_\infty\sim 1/M$ we see how an area law for the entropy results, $S_{22}\sim M^2$, due to the enormous volume $V_{22}$ associated with the conformally-related metric. This picture strictly only makes sense for massless particles, but there is a sense in which it also works for massive particles. This is because the actual temperature $T(x)$ can easily exceed any particle mass over most of the interior of a 2-2-hole, meaning that particle masses have little effect even when they are present. Their effect will show up as an edge effect near the surface of the 2-2-hole. In the large volume picture with constant temperature $T_\infty$, the same physics can be achieved by scaling particle masses smaller by the appropriate $x$-dependent conformal factor.

It is interesting that this large volume has recently been identified using thermodynamical arguments \cite{Aydemir:2021dan}, where it is referred to as the ``thermodynamic volume'' $V_{\rm th}$. This emerges in order to define the thermodynamics of a self-gravitating system in terms of intensive quantities measured at infinity, such as $T_\infty$ and $p_\infty$. If it was somehow possible to hold these quantities fixed, then a first law of thermodynamics could be considered $dU=T_\infty dS-p_\infty dV_{\rm th}$. The relations above show how this is satisfied as $\frac{3}{4}T_\infty dS_{22}=T_\infty dS_{22}-\frac{1}{4}T_\infty dS_{22}$. On the other hand such a variation does not relate different 2-2-hole solutions. We presently only have a one parameter family of solutions; the parameter can be taken to be $M$ and all other quantities vary with $M$.

Our previous numerical work \cite{Holdom:2019ouz} showed an especially simple dependence on $M$. We found that the relation $T_\infty S_{22}=M/2$ is either identically true or is very close to it. This would mirror the black hole relation $T_{\rm bh}S_{\rm bh}=M/2$. From (\ref{e15}) we would then have
\begin{align}
M=2T_\infty S_{22}=8p_\infty V_{22}=\frac{8}{3}U_{22}
.\end{align}
We shall see what our current approach says about the accuracy of $T_\infty S_{22}\approx M/2$, and as well we shall find the value of $\zeta$ where $T_\infty\approx T_{\rm bh}/\zeta$ and $S_{22}\approx\zeta S_{\rm bh}$.

We can also display the geodesic equations for a massless particle of energy $E$ and angular momentum $L$ moving in the $\theta=0$ plane,
\begin{align}
\frac{L^2}{E^2}\frac{1}{S(x)^2}=1-\dot{x}^2,\quad \frac{L}{E}\frac{1}{S(x)^2}=\dot{\phi}
.\label{e18}\end{align}
Conformal symmetry is again evident due to the absence of $R(x)$. The first equation implies $\ddot x=(L^2/E^2)S'(x)/S(x)^3<0$ and so the orbit reaches maximum $x_{\rm max}$ where $\dot x=0$ before falling back to the origin. It passes through the origin smoothly when viewed in polar coordinates; that is when $(x,\phi)$ is understood to mean $(-x,\phi+\pi)$ when $x<0$. When viewed using the conformally-related metric, the particle simply bounces off the last spherical surface at $x=0$.

An orbit has a value of $L/E=S(x_{\rm max})$ that can be enormous and so particles of the gas typically have very large $L/E$. $L/E$ must then be unusually small for a particle to venture beyond the minimum of $S(x)$ at the light-ring radius, and so $x_{\rm lr}$ marks the peak in an angular momentum barrier that very effectively traps particles inside. For a massive particle there is an additional term $(1/E^2)R(x)^2/S(x)^2$ appearing on the lhs of the first equation (\ref{e18}) (both $L$ and $E$ are defined with the mass $m$ factored out). Now $x_{\rm max}$ occurs for $L=0$ and an $E=R(x_{\rm max})/S(x_{\rm max})$ that is typically very small. This is because massive particles in the thermal gas essentially only exist at radii inside where $T(x)\gtrsim m$. Then massive particles do not directly escape, although they remain in thermal equilibrium with very light or massless particles that do. As for a black hole, this radiation is small because $T_\infty$ is small.

In Section \ref{s2} we consider the wave equation on the 2-2-hole background.

\section{Solving the field equations}\label{s4}
Now we can get back to our main task which starts with the field equations. They can be written as \cite{Lu:2015psa}
\begin{align}
m_{\rm Pl}^2(R_{\mu\nu}-\frac{1}{2}g_{\mu\nu}R)-2q^2B_{\mu\nu}=8\pi T_{\mu\nu},\quad\quad q\equiv \frac{m_{\rm Pl}}{m_G}
\label{e2}.\end{align}
The Bach term originates from the Weyl-squared term and is traceless. Since we also assume $T^\mu_\mu=0$, the trace of (\ref{e2}) is $R=0$. This can be taken to be one of the field equations and it can be written as
\begin{align}
\frac{S''(x) S(x)-S'(x)^{2}+1}{S(x)^2}=\frac{3 R''(x)}{R(x)}
\label{e5}.\end{align}
For the other equation we take the $xx$ component of (\ref{e2}). For our choice of metric this takes a relatively simple form (with units $m_{\rm Pl}=1$),
\begin{align}
\frac{3 R'(x)^{2} S(x)^{2}-2 R'(x) R(x) S'(x)S(x) -R(x)^{2}}{R(x)^{2} S(x)^{2}}&-q^2\frac{S'(x)^{4}+ S''(x)^{2}S(x)^{2}-2  S'''(x)S'(x) S(x)^{2}-1}{3 R(x)^{2} S(x)^{2}}\nonumber\\
&=8\pi\frac{R(x)^2}{S(x)^2}p(x)\label{e3}
.\end{align}
The absence of $R(x)$ derivatives in the second term is related to the conformal invariance of the Weyl-squared term. This equation is left with only first derivatives of $R(x)$.

We first consider no matter, $p(x)=0$, and the black hole solution. This is a solution with or without the Bach term, and so the two terms on the lhs of (\ref{e3}) each vanish separately. The vanishing of the second term thus determines $S(x)$, and we have already given the solution in (\ref{e4}). In (\ref{e5}) we note that the $ S''(x) S(x)-S'(x)^{2}+1\to3$ for $x\to-\infty$. Since $S(x)$ is then growing large, we see why $R''(x)$ is decreasingly small while positive. Also since $S'(x)$ is negative here, we see from the vanishing first term in (\ref{e3}) why $R'(x)$ is also decreasingly small and positive. With these behaviors for $R''(x)$ and $R'(x)$, $R(x)$ approaches a constant from above for decreasing $x$. This constant is $2M$.

Now we want to see how this black hole behavior transitions to the interior 2-2-hole solution. The latter is characterized by having $S(x)$ approach a large finite value at $x=0$, with a vanishing first derivative and a negative second derivative. Then $ S''(x) S(x)-S'(x)^{2}+1$ is large and negative at the origin. Thus in the transition region this quantity must change sign from positive to negative for decreasing $x$, thus forcing $R''(x)$ to do the same from (\ref{e5}). Meanwhile the second term in (\ref{e3}) no longer vanishes and the combination $S'(x)^{4}+ S''(x)^{2}S(x)^{2}-2  S^{\prime \prime \prime} (x)S'(x) S(x)^{2}$ grows larger than unity. The pressure term reinforces this effect in the equation.\footnote{This suggests that matter is actually not necessary for the solution. This is true since a 2-2-hole was first found numerically as a vacuum solution \cite{Holdom:2002xy}. We expect such a solution to be unstable.} Then the first term on the lhs must compensate and this forces $R'(x)$ to grow larger while remaining positive. With these behaviors for $R''(x)$ and $R'(x)$, $R(x)$ is forced smaller than $2M$ and we smoothly go over to the interior solution of the 2-2-hole. In the interior both terms on the lhs of (\ref{e3}) contribute leading terms that are $\sim 1/M^2$, since $R(x)\sim M$ (away from the origin), $S(x)\sim M^2$ and each derivative brings in a factor $1/M$. And since $p(x)\sim T(x)^4\sim (S(x)/R(x))^4T_\infty^4$, we see that $T_\infty\sim 1/M$.

With this rough understanding of the 2-2-hole solution, we can turn to a series expansion of the field equations around the origin $x=0$. We consider
\begin{align}
R(x)&=r_1x+r_3x^3+r_5x^5+...\nonumber\\
S(x)&=s_0(1+s_2x^2+s_4x^4+...)
.\end{align}
It is convenient to pull out the factor of $s_0$ as we will find that it scales as $M^2$, and then the other coefficients will scale, at leading order in $1/M$, as $s_i\sim M^{-i}$ for $i=2,4,...$ and $r_i\sim M^{1-i}$ for $i=1,3,...$. The following results will be further simplified if we write the rhs of (\ref{e3}) as 
\begin{align}
(3 - \frac{4}{3} c^2) \frac{r_1^2}{s_0^2} \frac{S(x)^2}{R(x)^2} 
,\label{e9}\end{align}
with a new constant $c<3/2$ that will be related to $T_\infty$.

After inserting this expansion into the field equations and solving for the coefficients, we find that all coefficients can be expressed in terms of $r_1$, $s_0$ and $c$. Furthermore, since only terms leading in $1/M$ are kept, it is observed that the dependence on $r_1$ and $s_0$ (and $q$) can be removed by a certain rescaling. In particular we can consider the rescaled functions 
\begin{align}
\hat R(\hat x)&=\frac{1}{\sqrt{r_1s_0q}}R(\frac{\sqrt{s_0q}}{\sqrt{r_1}}\hat x)\nonumber\\
\hat S(\hat x)&=\frac{1}{s_0}S(\frac{\sqrt{s_0q}}{\sqrt{r_1}}\hat x)
\label{e7},\end{align}
and the corresponding expansions
\begin{align}
\hat R(\hat x)&=\hat x+\hat r_3\hat x^3+\hat r_5\hat x^5+...\nonumber\\
\hat S(\hat x)&=1+\hat s_2\hat x^2+\hat s_4\hat x^4+...\;
.\label{e21}\end{align}
The coefficients of these expansions depend only on $c$, in particular
\begin{align}
\hat R(\hat x)&=\hat x- \frac{1}{9}c\hat x^3+(\frac{1}{10}-\frac{7}{54}c^2)\hat x^5+c(\frac{19}{180}-\frac{401}{3402} c^2)\hat x^7+...\nonumber\\
\hat S(\hat x)&=1- c\hat x^2+(\frac{1}{2}-\frac{1}{6} c^2)\hat x^4- c(\frac{1}{36}+\frac{1}{30} c^2)\hat x^6+...\;
.\label{e6}\end{align}

$\hat R(\hat x)$ and $\hat S(\hat x)$ are universal functions that determine the interior spacetime in the large $M$ limit. The corrections to this limit are suppressed by $1/M^2$ and are thus completely negligible for all but the smallest 2-2-holes. It is therefore useful to have the differential equations that determine $\hat R(\hat x)$ and $\hat S(\hat x)$ directly. These are obtained by using (\ref{e7}) to transform (\ref{e5}) and (\ref{e3}) into equations with $R\to\hat R$, $S\to\hat S$ and $x\to\hat x$. Two terms are found to be suppressed by $s_0^{-1}$ and a third by $s_0^{-2}$. After removing these terms then the resulting equations are
\begin{align}
\frac{\hat S''\hat S-\hat S'^{2}}{\hat S^2}&=\frac{3 \hat R''}{\hat R},\\
3 \hat R'^{2}\hat S^{2}-6 \hat R' \hat R\hat S'\hat S-\hat S'^{4}+\hat S''^{2}\hat S^{2}-2 \hat S'''\hat S' \hat S^{2}&=(9 - 4 c^2)\hat S^4
\label{e13}.\end{align}
Note that these equations cannot be used for the exterior or black hole solution, for that case the dropped terms are needed. A series expansion of these equations now gives (\ref{e6}) directly. The solutions as determined by $c$ are now straightforward to explore numerically. The integration cannot actually be started at $\hat x=0$, but the series expansion can be used to generate the initial conditions needed to start the integration at a finite but small $\hat x$. At this $\hat x$ the magnitude of the highest order term in the series expansion should be less than the precision used to specify the initial conditions. The accuracy of the final results is also determined by the precision of the numerical integration and by the procedure that is described in the next section.

We note that a high order series expansion can be used by itself to produce quite informative results with no need of numerical solutions. But we obtain more accurate results when using numerical integration in combination with the series expansion.

We can also briefly consider the full quadratic gravity theory in (\ref{e8}) and the effect of the $R^2$ term. For a finite $m_S$ the additional terms on the lhs of field equation (\ref{e2}) are
\begin{align}
\frac{m_{\rm Pl}^2}{3m_S^2}\left(R_{\mu\nu}-\frac{1}{4}g_{\mu\nu}R+g_{\mu\nu}\Box -\nabla_\mu\nabla_\nu\right)R
.\end{align}
Thus if $R$ is everywhere zero then these terms vanish, which means that our solutions are also solutions of the full theory. A similar situation was discussed in \cite{Lu:2015psa}. We have nevertheless carried out the series expansion of the full field equations following the steps above. The result in the form of (\ref{e21}) has all coefficients depending on both $c$ and $m_{\rm Pl}/m_S$, as well as on an undetermined $\hat s_2$. This additional parameter corresponds to the additional degree of freedom in the theory. If $\hat s_2$ is set to $-c$ then the expansion collapses back to (\ref{e6}). But it is possible that $\hat s_2$ and $c$ could be tuned simultaneously to find some other thermal-gas solution of the full theory with finite $m_S$. We do not consider this further.

\section{Determining the parameters}\label{s5}
The parameters that now need to be determined are $c$, the values of $r_1$ and $s_0$ to recover $R(x)$ and $S(x)$, and then $x_{\rm lr}$ and $T_\infty$. We will be able to determine these parameters from the numerical solutions and from our knowledge of how these functions need to approach the black hole solution for increasing $x$.

\begin{figure}[b]
\includegraphics[width=.8 \textwidth]{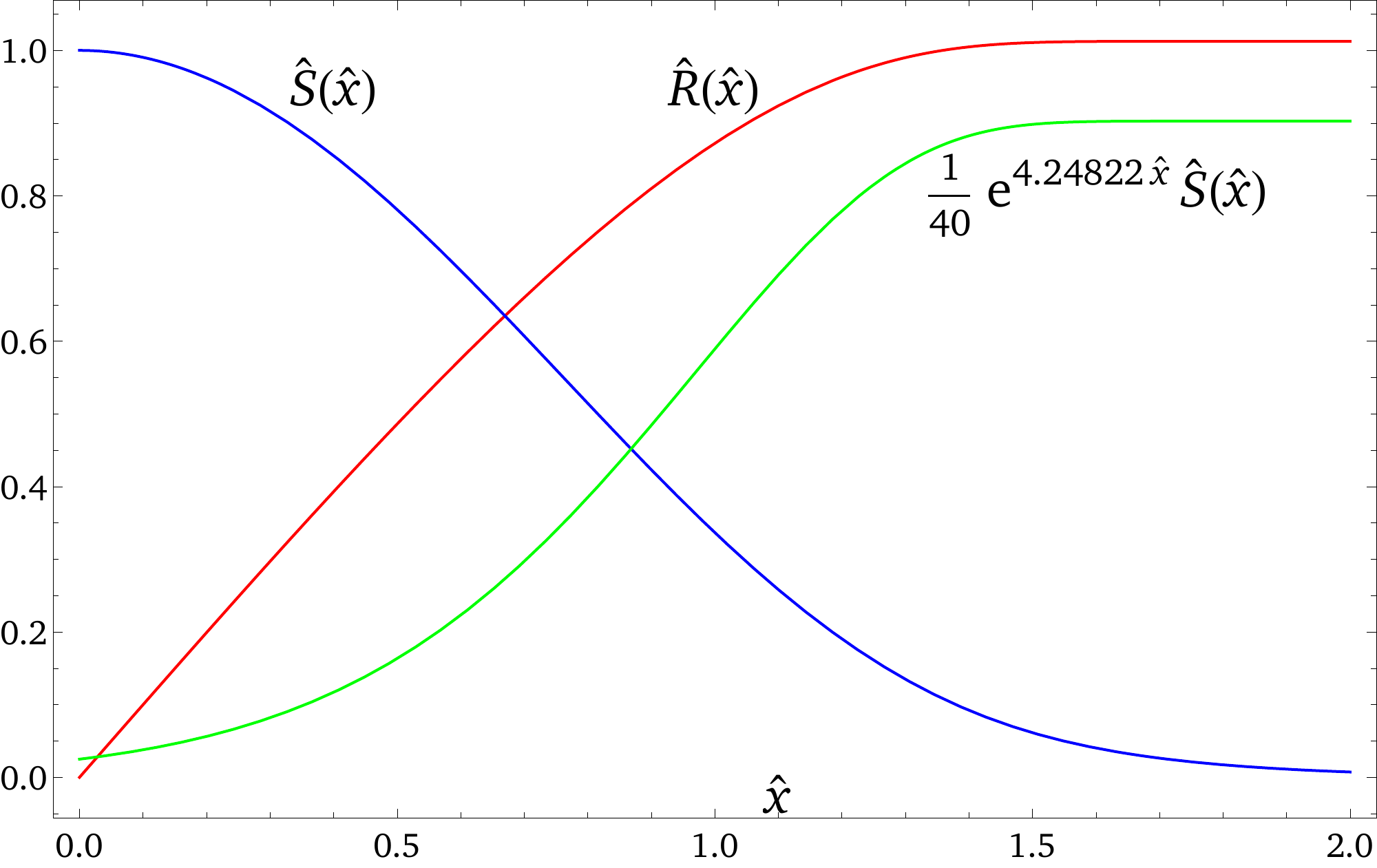}
\caption{$\hat R(\hat x)$ and $\hat S(\hat x)$ times an exponential function both approach constants from below.}\label{f2}
\end{figure}

Determine $c$. We show $\hat R(\hat x)$ in Fig.~\ref{f2} for a value of $c$ at which $\hat R(\hat x)$ smoothly and monotonically approaches a constant from below. The shooting method constrains $c$ to high accuracy as $c=0.95881880401$. In fact a much higher precision number than this is used, but its precise value depends on the details of the high precision numerical integration. 

Determine $r_1 s_0$. We find that $\hat R(\hat x)$ approaches the value $1.01250$ from below. But we know that $R(x)$ must approach $2M$ from below, and thus from the scaling in (\ref{e7}) we have $1.0125\sqrt{r_1 s_0q}\approx2M$.

Determine $s_0/r_1$. We also know that the 2-2-hole $S(x)$ solution must smoothly approach the black hole solution $S_{\rm bh}(x)$ from below. We see from (\ref{e11}) that $S_{\rm bh}(x)e^{x/(4M)}$ is a constant in this range, and thus $S(x)e^{x/(4M)}$ must approach a constant from below. Equivalently the same must hold for
\begin{align}
\hat S(\hat x)\exp\left(\frac{1}{4M}\frac{\sqrt{s_0q}}{\sqrt{r_1}}\hat x\right)
.\label{e10}\end{align}
We show this function in Fig.~\ref{f2}, and the desired behavior occurs for $\sqrt{s_0q/r_1}/(4M)\approx4.24822$. Thus with this result for $s_0/r_1$ and the previous one for $r_1 s_0$, we find $r_1\approx 0.1162$ and $s_0\approx 33.57 M^2/q$. As a reminder, these are the values of $R'(0)$ and $S(0)$ at the origin of a large 2-2-hole of mass $M$.

Determine $x_{\rm lr}$. Given $r_1$ and $s_0$ we can construct $S(x)$ from $\hat S(\hat x)$. For increasing $x$, $S(x)$ should eventually match the exponentially falling behavior of $S_{\rm bh}(x)$ in (\ref{e11}). This then determines the $x_{\rm lr}$ parameter appearing in $S_{\rm bh}(x)$ and the result is
\begin{align}
x_{\rm lr}(M)\approx4M\log(550M/q)
\label{e12}.\end{align}
We now have the information needed to generate Fig.~\ref{f1}, which we did for $q=1$. The range of the red curves in that figure is equivalent to the $0<\hat x<2$ range in Fig.~\ref{f2}.

The numerical coefficient of 4 in (\ref{e12}) is exact and it can be traced to the $1/4$ appearing in the exponential behavior of $S_{\rm bh}(x)$ in (\ref{e11}), since (\ref{e12}) can be written as $e^{x_{\rm lr}/(4M)}=aM$ with $a\approx550/q$. $S(x)$ is growing by a factor of $\sim M$, from $\sim M$ at $x=x_{\rm lr}$ to $\sim M^2$ at $x=0$. $S(x)$ deviates from exponential growth at the two ends of this range where it levels off to a vanishing slope. This leads to a larger range of $x$ for the same growth in $S(x)$ which in turn leads to the large value of $a$.

Determine $T_\infty$. We determine the constant $\zeta$ defined by $\zeta T_\infty=T_{\rm bh}=1/(8\pi M)$. Combining the rhs of (\ref{e3}), proportional to $\zeta^{-4}$, and (\ref{e9}) and inserting our values of $c$, $r_1$ and $s_0$ gives
\begin{align}
\zeta\approx 0.7548 N^\frac{1}{4}q^{-\frac{1}{2}}
.\label{e17}\end{align}

From (\ref{e15}) we see that $T_\infty S_{22}$ is also proportional to $\zeta^{-4}$. Thus independently of (\ref{e17}), we can ask for what $\zeta$ is $T_\infty S_{22}=T_{\rm bh}S_{\rm bl}=M/2$ true? Our numerical result for $S(x)$ allows us to calculate the integral in $V_{22}$ and $S_{22}$, and the $\zeta$ as determined in this way is the same as the one in (\ref{e17}) to better that $10^{-6}$ accuracy. This is strong evidence that $T_\infty S_{22}=M/2$ is an exact relation in the large $M$ limit. This result is striking, since the first value of $\zeta$ comes from solving the field equations, while the second comes from performing a certain integral over $S(x)$.

We can compare to previous results by converting our expansion parameters $r_1$ and $s_0$ into the expansion parameters appearing in $A(r)=a_2r^2+...$ and $B(r)=b_2r^2+...$. From $a_2=r_1^{-2}s_0^{-2}$ and $b_2=s_0^{-2}$ we have $a_2\approx0.0657q^2/M^4$ and $b_2\approx0.00089q^2/M^4$. These results are similar to previous results at smaller $M$, as is the result for $\zeta$ in (\ref{e17}) \cite{Holdom:2019ouz}.

Our present approach applies in the large $M$ limit, and so we may ask how large does $M$ have to be for the equations in (\ref{e13}) to be a good approximation to (\ref{e5}) and (\ref{e3}). We have dropped two terms of order $1/M^2$, and these dropped terms must be smaller in magnitude than the combinations 
$S'(x)^2-S''(x)S(x)$ and $(3R'(x)^2S(x)^2-2R'(x)S(x)R(x)S'(x))/R(x)^2$ respectively. These quantities smoothly drop for increasing $\hat x$, and become $\sim 10^{-8}$ for $\hat x=2$. Thus if we want to use our results for $\hat R(\hat x)$ and $\hat S(\hat x)$ throughout the range $0<\hat x<2$, we need $M\gtrsim10^5$. This motivated the choice $M=10^5$ for Fig.~\ref{f1}. From that figure we can see that the range $0<\hat x<2$ corresponds to some fraction of the range $0<x<x_{\rm lr}$. For general $M$ we find this fraction to be $\approx8.5/\log(550M/q)$. Thus the relative horizontal extent of the red curves in Fig.~\ref{f1} is corresponding smaller for much larger $M$.

\section{Gravitational waves}\label{s2}
Consider some perturbation of a field $\psi$ with spin $s$ in a general spherically-symmetric static spacetime. $s=0,1,2$ corresponds to massless scalar, electromagnetic and gravitational fields, where for the last case we are referring to the Regge-Wheeler equation. We take the decomposition into the angular momentum modes $\Psi_l(x)$ as $\psi=\sum_{lm} e^{-i\omega t}\frac{1}{R(x)}\Psi_l(x) Y_{lm}(\phi,\theta)$, where the $1/R(x)$ in this definition removes a $R^\prime(x)\Psi_l^\prime(x)$ term in the radial wave equation for $\Psi_l(x)$. This radial equation is then
\begin{align}
&\left(\partial_x^2+\omega^2-V_{s,l}(x)\right)\Psi_l(x)=0,\\
&V_{s,l}(x)\equiv V_R(x)+V_S(x),\\
&V_R(x)=(1-s^2)\frac{R''(x)}{R(x)},\quad V_S(x)=\frac{l(l+1)}{S(x)^2}
\label{e1}.\end{align}
$S(x)$ appears in the angular momentum part of the potential. The $R(x)$ dependence is only absent in the $s=1$ case, corresponding to the fact that only the electromagnetic wave equation is conformally invariant.

Since $S(x)$ is very large in the interior of a 2-2-hole, $V_S(x)$ is typically negligible and no angular momentum barrier near the origin exists. $V_R(x)$ is also regular and so the radial wave for any $l$ satisfies a standard $s$-wave, Neumann boundary condition at the origin $x=0$. But it is $\Psi_l(x)/R(x)$ that satisfies this boundary condition and therefore $\Psi_l(x)$ satisfies a Dirichlet boundary condition. We shall find rather narrow ranges of $x$ where the total potential is non-negligible, and outside these ranges we basically just have plane-wave-like behavior for the radial wave propagation.

\begin{figure}[t]
\includegraphics[width=.8 \textwidth]{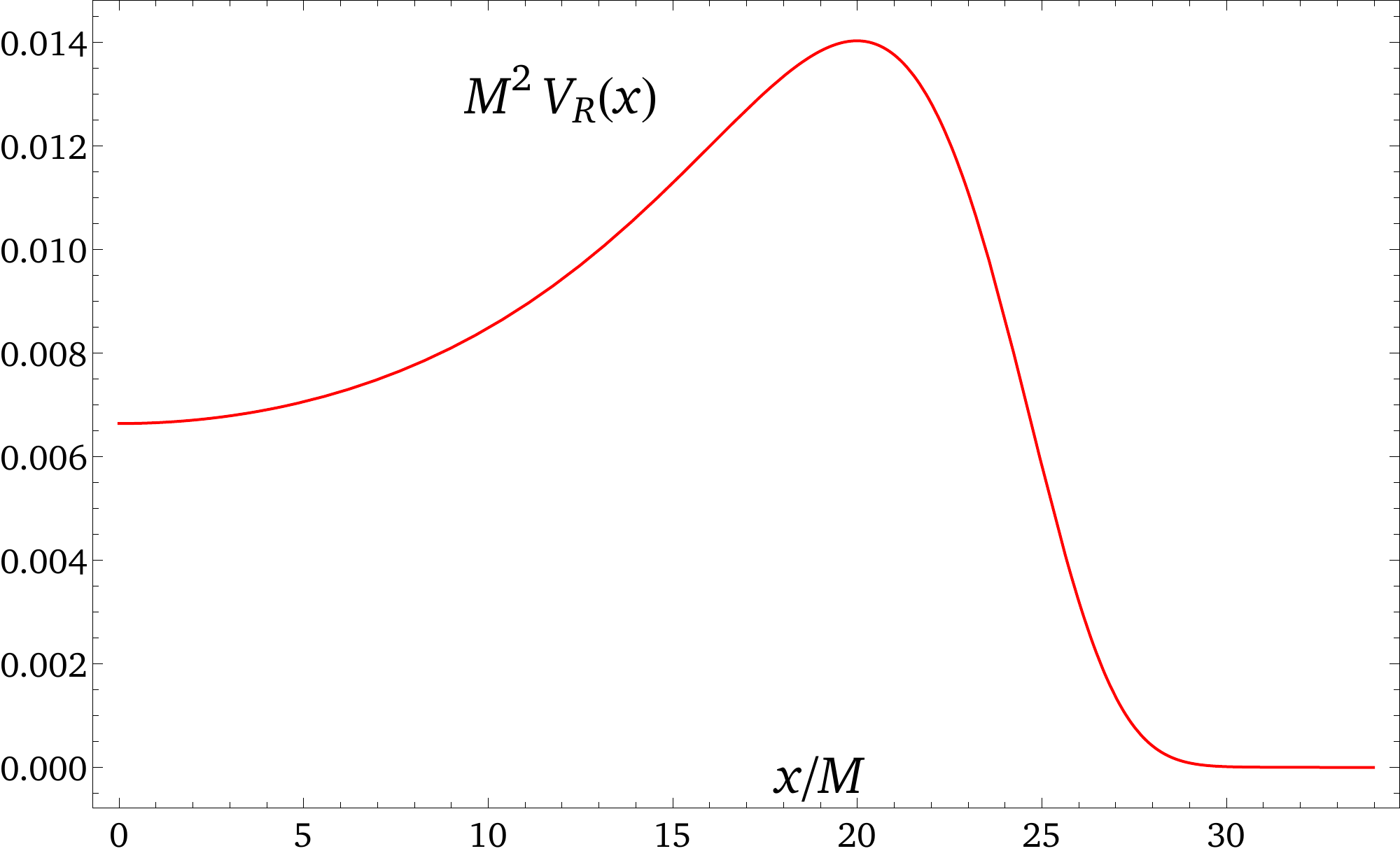}
\caption{The universal form of $V_R(x)$ with $s=2$ after scaling with $M$. The range of $x$ corresponds to $0<\hat x<2$.}\label{f3}
\end{figure}

For the Regge-Wheeler gravitational wave equation potential, $V_S(x)$ is specified with $s=2$. The contribution from $V_S(x)$ in the interior of a 2-2-hole is completely negligible, and so we display $V_R(x)$ in the interior in Fig.~\ref{f3}. This figure shows that $V_R(x)$ has a maximum value of $\approx0.014/M^2$.\footnote{$V_R(x)$ for a scalar wave equation has the opposite sign and it was displayed in \cite{Holdom:2016nek}.} This can be compared to the maximum value of $V_S(x)$ that occurs at the minimum value of $S(x)$, that is at $x=x_{\rm lr}$. This potential barrier height is $l(l+1)/(27M^2)$ where $l\geq s=2$. $V_R(x)$ is negative in this region and has a minimum value of $-81/(1024M^2)$ at $x=x_{\rm lr}-(2\log(3/2)+1/3)M$, but $V_S(x)$ dominates. Thus the potential generated in the interior of the 2-2-hole is smaller, but not negligible, compared to the standard black hole potential barrier around the light ring. This small interior potential is the only impact that the large curvatures have on the wave equation. In addition the wavelengths of interest are often comparable or larger than the radial extent of the interior. The gravitational wave can also interact with the matter in the interior as discussed in \cite{Holdom:2020onl}.

The surface of a 2-2-hole can be roughly defined to be the $x$ where the potential $V_R(x)$ in Fig.~\ref{f3} rises from zero. The $x$ where $R(x)=2M$ is larger, but this $x$ is not precisely determined in our approach. The $x$ where $R''(x)=0$ is larger still, and it is also not precisely determined. Another quantity with behavior similar to the potential $V_R(x)$ is the Weyl term in the action, evaluated on the 2-2-hole solution,
\begin{align}
\frac{\sqrt{-g}}{\sin(\theta)}C^{\mu\nu\alpha\beta}C_{\mu\nu\alpha\beta}=\frac{4}{3}\frac{(S''(x) S(x)-S'(x)^{2}+1)^2}{S(x)^2}=12\left(S(x)\frac{R''(x)}{R(x)}\right)^2
.\end{align}
(\ref{e5}) has been used for the last equality.

\section{Weak gravity conjecture}\label{s3}
There is another type of relativistic, conformally-invariant gas ($T_\mu^\mu=0$) that can act as the matter source for a 2-2-hole. And that is a cold ($T(x)=0$), degenerate, Fermi gas characterized by a local Fermi momentum $k_F(x)$. This quantity replaces the temperature of a thermal gas, and in a similar way, $k_F(x)=k_{F\infty} /\sqrt{-g_{tt}}$. $k_F(x)$ can again greatly exceed the fermion mass over most of the interior of the 2-2-hole. The local number density $n(x)$ replaces the entropy density as the quantity of interest. For a single species of fermion we have
\begin{align}
n(x)=\frac{1}{3\pi^2}k_F(x)^3,\quad \rho(x)=\frac{3}{4}k_F(x)n(x)=3p(x)
.\end{align}
We then get the total fermion number and the total fermion energy,
\begin{align}
N_{22}=\frac{1}{3\pi^2}k_{F\infty}^3 V_{22},\quad U_{22}=\frac{3}{4}k_{F\infty} N_{22}=3p_\infty V_{22}
,\label{e20}\end{align}
expressed in terms of the conformally-related volume as before. The similarity of the degenerate gas to the thermal gas means that we immediately have the results $N_{22}k_{F\infty}\approx M/2$, $U_{22}\approx 3M/8$ and $k_{F\infty}\approx T_{\rm bh}/\hat\zeta$ with $\hat\zeta\approx.40q^{-\frac{1}{2}}$.

A UV complete theory based on quadratic gravity is a quantum field theory. So let us entertain the possibility that such a theory can support the existence of a global or gauged $U(1)$ symmetry that does not obey the weak gravity conjecture. We assume that fermions carry the associated charge. Then 2-2-holes can carry a net charge of this type, and for sufficiently massive fermions the trapping of these fermions can be complete. Meanwhile the 2-2-hole can still radiate other degrees of freedom and thus continue to shrink and heat up. As this process continues, the trapped fermions make up an increasing fraction of the particle content of the 2-2-hole. Eventually the energy lost to radiation will lead to a decreasing temperature of the trapped fermions, and a cooling rather than a heating will occur as the 2-2-hole continues to shrink. The end point of this cooling is a degenerate Fermi gas, as we have just described.

Thus for a given global charge $N_{22}$, the mass and other properties of this cold and stable 2-2-hole are determined. Its entropy vanishes in the degenerate limit; all the information has escaped. There seems to be nothing conceptually peculiar about this object. This is quite unlike the analogous situation for a black hole, where arbitrary amounts of trapped charge implies an uncontrolled number of micro-states. This issue for black holes is used to motivate the weak gravity conjecture.

\section{Conclusion}
Our choice of metric cleanly separates those effects that arise due to the breaking of conformal symmetry through dependence on $R(x)$, and those that arise from the length scale $S(x)$. Many of our results are traced to the large size of $S(x)$ in a 2-2-hole, and this in turn is related to the exponential growth of $S(x)$ already seen in the black hole metric.

This exponential growth implies that the delay time between gravitational wave echoes has a $M\log M$ dependence with the same prefactor as for a truncated black hole model, when the truncation occurs at a proper Planck length from the would-be horizon. It appears that this result should continue to hold in general quadratic gravity, that is when the $R^2$ term is included. On the other hand we have also described a conformal-breaking contribution to the gravitational-wave-equation potential arising in the interior of the 2-2-hole. This is a feature that the truncated black hole model does not have, and this modified potential will affect the gravitational wave echo signal. We should stress though that spinning 2-2-hole solutions are needed to fully make contact with the realistic situation, and those solutions have not yet been found.

We discussed the calculation of the 2-2-hole entropy $S_{22}$ and found that it satisfies the relation $S_{22}T_\infty=M/2$. Compared to the black hole $S_{22}\approx \zeta S_{\rm bh}$ with $\zeta$ as in (\ref{e17}). For the species number $N\gtrsim 120$ as for the standard model and $q\equiv m_{\rm Pl}/m_G\approx1$, $S_{22}$ can easily be larger than $S_{\rm bh}$. Then a 2-2-hole cannot convert into the same mass black hole, assuming that the total entropy of a 2-2-hole and its surroundings cannot decrease and that energy is conserved. We can also compare the 2-2-hole entropy to the entropy $S_{\rm box}$ of a relativistic gas in a box of size $L$ in Planck units. It can be shown that $S_{\rm box}\lesssim N^\frac{1}{4}L^\frac{3}{2}$ to avoid gravitational collapse. This is smaller than the 2-2-hole entropy $S_{22}\sim N^\frac{1}{4}L^2$, but it is not necessarily smaller than black hole entropy $S_{\rm bh}\sim L^2$ for large $N$. In this way a 2-2-hole avoids the black hole ``species problem''.

We end by commenting on the conformally-related volume $V_{22}$ appearing in (\ref{e15}). To get a better sense of the size of this volume, consider some outside volume enclosed between $x_{\rm lr}$ and some much larger $x_>$. At what $x_>$ will this outside volume be the same as $V_{22}$? Numerically we find that $x_>\approx 33q^{-\frac{2}{3}} M^\frac{5}{3}$ in Planck units. This implies for example that the $x_>$ for a 2-2-hole of around 0.02 solar masses is similar to the size the observable universe. Thus one might say that the universe serves to glue together the attached 2-2-holes, where the latter have an effective total volume enormously larger than the universe itself.



\begin{thebibliography}{99}

\bibitem{Holdom:2016nek} 
  B.~Holdom and J.~Ren,
  ``Not quite a black hole'',
  Phys.\ Rev.\ D {\bf 95}, no. 8, 084034 (2017).
  [arXiv:1612.04889 [gr-qc]].

\bibitem{Holdom:2019ouz} 
  B.~Holdom,
  ``A ghost and a naked singularity; facing our demons,''
  arXiv:1905.08849 [gr-qc].
  
\bibitem{Ren:2019jft} 
  J.~Ren,
  ``Anatomy of a thermal black hole mimicker,''
  Phys.\ Rev.\ D {\bf 100}, no. 12, 124012 (2019)
  [arXiv:1905.09973 [gr-qc]].
  
\bibitem{Donoghue:2019ecz}
J.~F.~Donoghue and G.~Menezes,
Arrow of causality and quantum gravity,
Phys. Rev. Lett. \textbf{123}, 171601 (2019).
[arXiv:1908.04170 [hep-th]].

\bibitem{Donoghue:2019fcb}
J.~F.~Donoghue and G.~Menezes,
Unitarity, stability and loops of unstable ghosts,
Phys. Rev. D  \textbf{100}, 105006 (2019).
[arXiv:1908.02416 [hep-th]].

\bibitem{Holdom:2021hlo}
B.~Holdom,
``Ultra-Planckian scattering from a QFT for gravity,''
Phys. Rev. D \textbf{105}, 046008 (2022)
[arXiv:2107.01727 [hep-th]].

\bibitem{Holdom:2021oii}
B.~Holdom,
``Photon-photon scattering from a UV-complete gravity QFT,''
JHEP 04 (2022) 133,
[arXiv:2110.02246 [hep-ph]].

\bibitem{Lu:2015psa}
H.~L\"u, A.~Perkins, C.~N.~Pope and K.~S.~Stelle,
``Spherically Symmetric Solutions in Higher-Derivative Gravity,''
Phys. Rev. D \textbf{92}, no.12, 124019 (2015)
[arXiv:1508.00010 [hep-th]].

\bibitem{Svarc:2018coe}
R.~Svarc, J.~Podolsky, V.~Pravda and A.~Pravdova,
``Exact black holes in quadratic gravity with any cosmological constant,''
Phys. Rev. Lett. \textbf{121}, no.23, 231104 (2018)
[arXiv:1806.09516 [gr-qc]].
  
\bibitem{Boonserm:2008zg}
P.~Boonserm and M.~Visser,
``Bounding the greybody factors for Schwarzschild black holes,''
Phys. Rev. D \textbf{78}, 101502 (2008)
[arXiv:0806.2209 [gr-qc]].

\bibitem{Aydemir:2021dan}
U.~Aydemir and J.~Ren,
``On thermodynamics of compact objects,''
[arXiv:2201.00025 [gr-qc]].

\bibitem{Holdom:2002xy}
B.~Holdom,
``On the fate of singularities and horizons in higher derivative gravity,''
Phys. Rev. D \textbf{66}, 084010 (2002)
[arXiv:hep-th/0206219 [hep-th]].

\bibitem{Holdom:2020onl}
B.~Holdom,
``Damping of gravitational waves in 2-2-holes,''
Phys. Lett. B \textbf{813}, 136023 (2021)
[arXiv:2004.11285 [gr-qc]].


\end{thebibliography}
\end{document}